\documentclass[conference,a4paper]{IEEEtran}

\usepackage[utf8]{inputenc} 
\usepackage{graphicx}
\usepackage[T1]{fontenc}
\usepackage{url}
\usepackage{enumitem}
\usepackage[cmex10]{amsmath}
\usepackage{amssymb}
\usepackage{cite}
\usepackage{amsthm}
\usepackage{textcomp}
\usepackage{siunitx}
\usepackage{color}
\usepackage{subfigure}
\usepackage{cases}
\usepackage[font={small}]{caption}
\usepackage{tcolorbox}
\usepackage{fix-cm}

\makeatletter
\def\thm@space@setup{\thm@preskip=2pt
\thm@postskip=2pt \itshape}
\makeatother
\newtheoremstyle{newstyle}      
{} 
{} 
{\mdseries} 
{} 
{\bfseries} 
{.} 
{ } 
{} 

\theoremstyle{newstyle}

\newtheorem{theorem}{Theorem}

\theoremstyle{definition}

\newtheorem{definition}{Definition}

\theoremstyle{remark}
\newtheorem{remark}{Remark}

\setlist[description]{style=multiline}

\IEEEoverridecommandlockouts

\begin{document}
\sloppy

\setlength{\abovedisplayskip}{1mm}
\setlength{\belowdisplayskip}{1mm}
\setlength{\abovecaptionskip}{1mm}
\setlength{\belowcaptionskip}{-6pt}

\title{Compressed Coded Distributed Computing} 
\author{Songze~Li$^{*}$, Mohammad~Ali~Maddah-Ali $^{\dagger}$, and A.~Salman~Avestimehr$^{*}$\\
$^{*}$ Department of Electrical Engineering, University of Southern California, Los Angeles, CA, USA \\ 
$^{\dagger}$ Nokia Bell Labs, Holmdel, NJ, USA
}

\maketitle

\begin{abstract}
Communication overhead is one of the major performance bottlenecks in large-scale distributed computing systems, in particular for machine learning applications. 
Conventionally, compression techniques are used to reduce the load of communication by combining intermediate results of \emph{the same} computation task as much as possible.
 Recently, via the development of coded distributed computing (CDC), it has been shown that it is possible to enable coding opportunities across intermediate results of \emph{different} computation tasks to further reduce the communication load.
 We propose a new scheme, named \emph{compressed coded distributed computing} (in short, \emph{compressed CDC}), which jointly exploits the above two techniques (i.e., combining the intermediate results of the same computation and coding across the intermediate results of different computations) to significantly reduce the communication load for computations with linear aggregation (reduction) of intermediate results in the final stage that are prevalent in machine learning (e.g., distributed training algorithms where partial gradients are computed distributedly and then averaged in the final stage).
 %
In particular, compressed CDC first compresses/combines several intermediate results for a single computation, and then utilizes multiple such combined packets to create a coded multicast packet that is simultaneously useful for multiple computations.
 We characterize the achievable communication load of compressed CDC and show that it substantially outperforms both combining methods and CDC scheme.

\end{abstract}

\section{Introduction}
In order to scale up machine learning applications that process a massive amount of data, various distributed computing frameworks have been developed where data is stored and processed distributedly on multiple cores or GPUs on a single machine, or multiple  machines in computing clusters (see, e.g., \cite{dean2004mapreduce,zaharia2010spark,recht2011hogwild}). When implementing these frameworks, the communication overhead of shuffling intermediate results across distributed computing nodes is a major performance bottleneck. For example, it was observed in~\cite{chowdhury2011managing} that on a Facebook's Hadoop cluster, 33\% of the job execution time was spent on data shuffling. This bottleneck is becoming worse for training deep neural networks with millions of model parameters (e.g., ResNet-50~\cite{he2016deep}) using distributed stochastic gradient descent, where partial gradients with millions of entries need to be passed between computing nodes.

Conventionally, compression techniques are used to reduce the communication load by combining intermediate results of \emph{the same} computation task as much as possible. For example, in the original MapReduce distributed computing framework~\cite{dean2004mapreduce}, when the Reduce function is commutative and associative, a ``combiner function'' is proposed to pre-combine multiple intermediate values with the same key computed from different Map functions. Then, instead of sending multiple values to the reducer, the mapper only needs to send the pre-combined value whose size is the same as one of the values before combining, which significantly reduces the bandwidth consumption without any performance loss.


Coded distributed computing (CDC) is another approach that has been recently proposed in~\cite{LMA_all,li2016fundamental}  to mitigate the communication bottleneck. 
Unlike the compression/combining technique, CDC enables coding opportunities across intermediate results of \emph{different} computation tasks to further reduce the communication load. In particular, within a MapReduce-type distributed computing model, CDC specifies a repetitive pattern of computing Map functions, creating side information at the computing nodes that enables coded multicasting during data shuffling across nodes, where each coded multicast packet is simultaneously useful for multiple Reduce tasks. For example, if we repeat each of the Map tasks $r$ times across the cluster, then utilizing the CDC scheme, we can reduce the total amount of bandwidth consumption by $r$ times. It has been shown that CDC can provide substantial speedups in practice~\cite{li2017CTS}, and several generalizations of it has been developed in the literature~\cite{li2016scalable,LMA-Allerton16,ezzeldin2017communication,kiamari2017heterogeneous,konstantinidis2018leveraging}. 


In this paper, we focus on MapReduce-type distributed computing frameworks and propose a new scheme, named \emph{compressed coded distributed computing} (in short, \emph{compressed CDC}). It jointly exploits the above compression/combining technique and the CDC scheme to significantly reduce the communication load for computation tasks with linear Reduce functions (and arbitrary Map functions) that are prevalent in data analytics (e.g., distributed gradient descent where the partial gradients computed at multiple distributed computing nodes are averaged to reduce to the final gradient).
Specifically, 
the compressed CDC scheme first specifies a repetitive storage of the dataset across distributed computing nodes. Each node, after processing locally stored files, first pre-combines the intermediate values of a single computation task needed by another node. Having generated multiple such pre-combined packets for different tasks, the computing node further codes them to generate a coded multicast packet that is simultaneously useful for multiple tasks. Therefore, compressed CDC enjoys both the intra-computation gain from combining, and the inter-computation gain from coded multicasting.




We characterize the achievable communication load of compressed CDC and show that it substantially outperforms both combining methods and CDC scheme. In particular, compared with the scheme that only relies on the combining technique, compressed CDC reduces the communication load by a factor that is proportional to the storage size of each computing node, which is significant for the common scenarios where large-scale machine learning tasks are executed on commodity servers with relatively small storage size.
On the other hand, compared with the CDC scheme whose communication load scales linearly with the size of the dataset, compressed CDC eliminates this dependency by pre-combining intermediate values of the same task, allowing the system to scale up to handle computations on arbitrarily large dataset. 


\subsection*{Other Related Work}
Motivated by the fact that training algorithms exhibit tolerance to precision loss of intermediate values, as opposed to the above lossless compression technique that guarantees exact recovery of computation results, a family of lossy compression (or quantization) algorithms for distributed learning systems have been developed to compress the intermediate results (e.g., gradients) for a smaller bandwidth consumption
(see, e.g.,~\cite{seide20141,alistarh2017qsgd,wen2017terngrad}). Apart from compression, various coding techniques have also been recently utilized in distributed machine learning  algorithms to mitigate the communication bottleneck and the straggler's delay
(see, e.g.~\cite{lee2015speeding,dutta2016short,TLDK16,LMA16_unify,li2017codingfog,yu2017polynomial,yu2018straggler,song2017pliable,attia2016information}).
\section{Motivating Example}\label{sec:examples}
In this section, we demonstrate through a motivating example, how compression and CDC techniques, applied alone or jointly, can help to reduce the bandwidth requirement for distributed computing tasks.

\begin{figure}[htbp]
  \centering
  \includegraphics[width=0.4\textwidth]{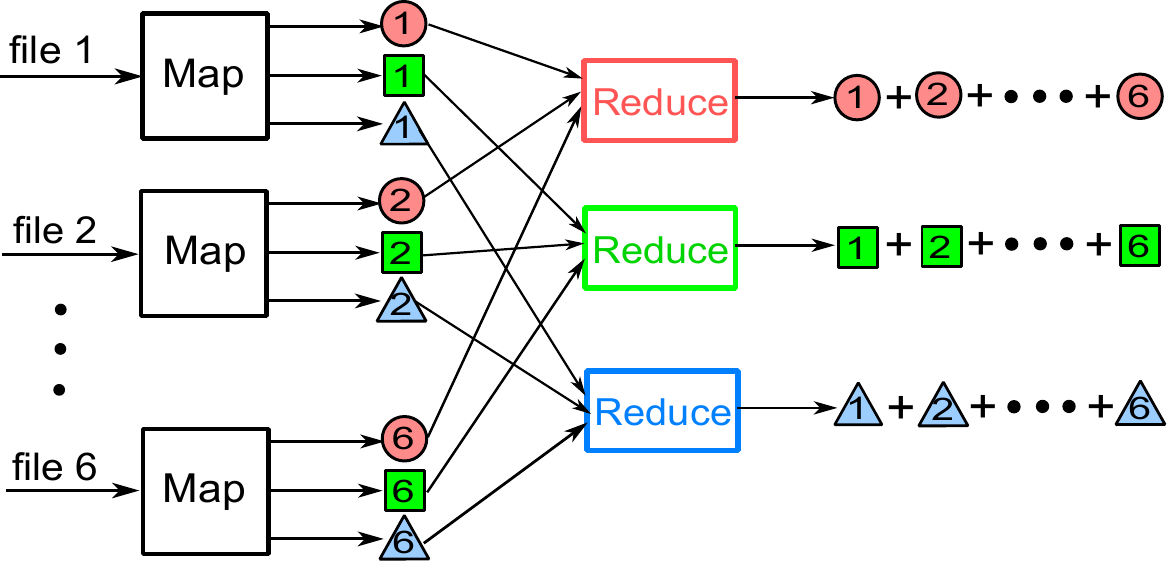}
  \caption{A MapReduce framework to compute 3 functions from 6 files with linear Reduce functions.}
  \label{fig:framework}
\end{figure}

\begin{figure*}[htbp]
  \centering
  \includegraphics[width=1\textwidth]{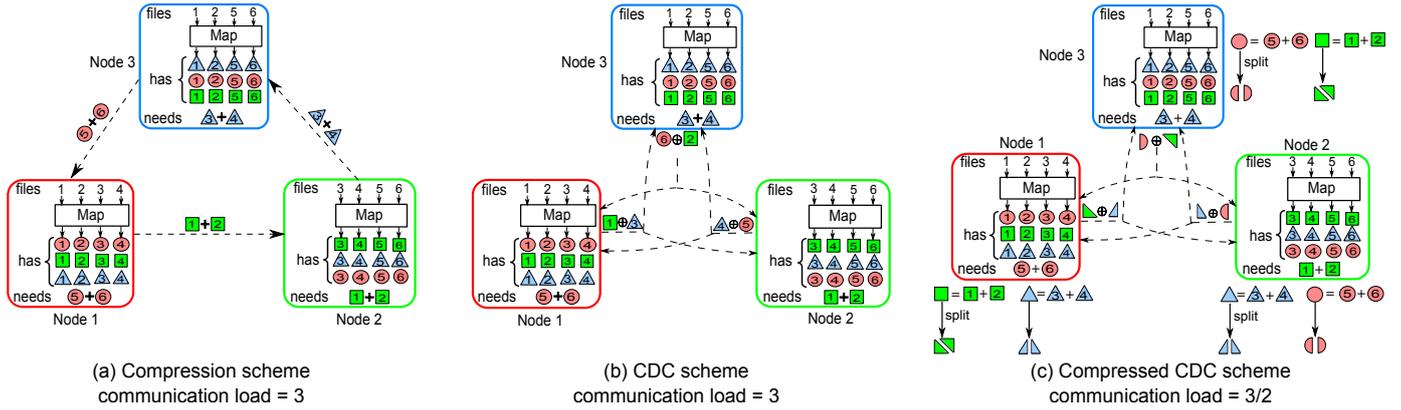}
  \caption{Coded computing schemes for a MapReduce job with linear Reduce functions, which processes $6$ files to compute $3$ functions, over $3$ distributed computing nodes each with a storage size of $4$ files.}
  \label{fig:motivating-example}
\end{figure*}

As shown in Fig.~\ref{fig:framework}, we consider a MapReduce job of computing $3$ output functions, represented by red/circle, green/square, and blue/triangle respectively, by processing $6$ input files. When mapping a file, we obtain $3$ intermediate values, one for each of the functions, which are represented by the color/shape of the corresponding functions labelled by the file index. The Reduce operation of each output function computes its final result by summing up the intermediate values of the function from all $6$ input files. This computation job is executed on $3$ distributed computing nodes connected through a multicast network. Each node can store up to $4$ files in its local memory. As shown in Fig.~\ref{fig:motivating-example}, we assign the computation tasks such that Nodes 1, 2, and 3 are respectively
responsible for final reduction of red/circle, green/square, and
blue/triangle functions. For this problem, we are interested in minimizing the communication load, which is the number of bits that need to be shuffled between computing nodes to accomplish the computation tasks, normalized by the size of a single intermediate value. Next, we describe three coded computing schemes, and compare their communication loads.




For all of these three schemes, as illustrated in Fig.~\ref{fig:motivating-example}, 
the file placement is performed such that Node~1 stores the files $1,2,3,4$, Node~2 stores the files $3,4,5,6$, and Node~3 stores the files $5,6,1,2$.

\subsubsection{Compression scheme}


Since only the sum of the intermediate values is needed for final reduction, we can pre-combine the computed intermediate values of the same function at the sender node to reduce communication. For example, as shown in Fig.~\ref{fig:motivating-example}(a), having computed the green squares labelled by $1$ and $2$ in the Map phase, Node~1 sums them up and sends the computed sum to Node~2, instead of sending them individually. Upon receiving this pre-combined packet, Node~2 can directly use it for the final reduction of the green/square function. This compression scheme reduces the communication load by half, compared with the schemes that unicast uncoded intermediate values, and achieves a communication load of $3$. 





\subsubsection{CDC scheme} Utilizing the redundant Map results across computing nodes, the CDC scheme creates coded multicast packets by combining intermediate values of different functions that are intended at different nodes. 
As shown in Fig.~\ref{fig:motivating-example}(b), since the blue triangle labelled by 3 is computed at both Nodes 1 and 2, and the green square labelled by 1 is computed at both Nodes 1 and 3, Node 1 can multicast the bit-wise XOR (denoted by $\oplus$) of these two intermediate values to the other two nodes. From this coded packet, both Nodes~2 and 3 can decode their intended values by cancelling their locally computed values. 
Since each of the multicast packets is simultaneously useful for two nodes, the CDC scheme cuts the communication load by half from the schemes that unicast uncoded intermediate values, and achieve a communication load of $3$. While achieving the same communication load as the compression scheme that pre-combines intermediate values of the same function, the CDC scheme combines intermediate values from different functions, and allows the recovery of them individually instead of their sum. Therefore, CDC can be utilized on more general MapReduce jobs with arbitrary Reduce functions to slash the communication load.

\subsubsection{Compressed CDC scheme} The above described two techniques can be applied jointly to further reduce the communication load. In particular, we can generate coded multicast packets as in the CDC scheme from the pre-combined packets created as in the compression scheme. Each node, as shown in Fig.~\ref{fig:motivating-example}(c), sums up two pairs of intermediate values to generate two pre-combined packets, each of which is needed by another node. Then, for example, Node~1 first splits each of its pre-combined packets (the unlabelled green square and the unlabelled blue triangle) into two segments, and computes the bitwise-XOR, of two segments, one from each of the pre-combined packets, generating a coded packet whose size is half of the size of an intermediate value. Finally, Node~1 multicasts this coded packet to Nodes 2 and 3. Similar operations are performed at Nodes 2 and 3. Next, each node utilizes the locally computed intermediate values to decode the intended pre-combined packet, which is used to reduce the output function. Compared with the compression and the CDC schemes, the compressed CDC scheme exploits both the compression opportunities within individual functions, and the multicasting opportunities across different functions, and achieves a communication load of $\frac{3}{2}$. 

In the next section, we first give the general problem formulation, and then present our main results on the proposed coded computing scheme that jointly exploits both types of coding from the compression scheme and the CDC scheme.

\section{Problem Formulation and Main Results}
We consider a computation job of processing $N$ input files, for some $N \in \mathbb{N}$, to compute $Q$ output functions, for some $Q \in \mathbb{N}$. We denote the $N$ input files as $w_1,\ldots,w_N \in \mathbb{F}_{2^F}$, for some $F \in \mathbb{N}$, and the $Q$ output functions as $\phi_1,\ldots,\phi_Q: (\mathbb{F}_{2^F})^N \rightarrow \mathbb{F}_{2^T}$, for some $T \in \mathbb{N}$. We focus on a class of computation jobs with \emph{linear} aggregation for which the computation of each output function can be decomposed as the sum of $N$ intermediate values computed from the input files, i.e., for $q = 1,\ldots,Q$,
\begin{align}
    \phi_q(w_1,\ldots,w_N) =  v_{q,1}+ v_{q,2} \cdots+v_{q,N}, \label{eq:linearReduce}
\end{align}
where $v_{q,n} = g_q(w_n)$ is the intermediate value of $\phi_q$ computed from some intermediate function $g_q: \mathbb{F}_{2^F} \rightarrow \mathbb{F}_{2^T}$. 
So far, we have introduced one computation job that involves computing $Q$ functions. Here, we consider the scenario where $J$ such computation jobs are executed in parallel, for some $J\in \mathbb{N}$. We denote the $N$ input files of job $j$ as $w_{1^{(j)}},\ldots,w_{N^{(j)}}$, and the $Q$ output functions job $j$ wants to compute as $\phi_{1^{(j)}},\ldots,\phi_{Q^{(j)}}$.\footnote{As an example, we can consider executing $J$ machine learning tasks (e.g., image classification), each of which has its own dataset, and aims to obtain its own set of model parameters. 
Another example is the navigation application, where $J$ navigation sessions, each of which requires to find the shortest path on a disjoint sector of the map, are executed in parallel.}


\subsection{Network model}
The above described $J$ computation jobs are executed distributedly on a computer cluster that consists of $K$ distributed computing nodes, for some $K \in \mathbb{N}$. These computing nodes are denoted as Node~$1,\ldots,$ Node~$K$. Here we assume $K \leq N$, and focus on a symmetric setting for the sake of load balancing, in which $K | Q$, and each node is responsible for computing $\frac{Q}{K}$ output functions for each job. The $K$ nodes are connected through an error-free broadcast network. Each node has a local storage that can store up to $\mu J N$ input files, i.e., $\mu$ fraction of the entire dataset that contains all input files from all jobs, for some $\mu$ satisfying $\frac{1}{K} \leq \mu < 1$. 


Before the computation starts, each node selects and stores $\mu J N$ input files from the dataset. For each node $k$, we denote the set of indices of the files stored locally as ${\cal M}_k$. 
A valid file placement has to satisfy 1) $|{\cal M}_k| \leq \mu J N$, for all $k=1,2,\ldots,K$ (local storage constraint), and 2) $\cup_{k=1,\ldots,K}{\cal M}_k = \cup_{j=1,\ldots,J}\{n^{(j)}:n=1,2,\ldots,N\}$ (the entire dataset needs to be collectively stored across the cluster).
\subsection{Distributed computing model}
The $K$ nodes process their locally stored files to compute the output functions following a MapReduce-type model. In particular, the overall computation proceeds in three phases: Map phase, Shuffle phase, and Reduce phase. 

\noindent {\bf Map phase.} For each file $w_{n^{(j)}}$ of job $j$, $n^{(j)} \in {\cal M}_k$, Node~$k$ maps it into $Q$ intermediate values $v_{1^{(j)},n^{(j)}}, v_{2^{(j)},n^{(j)}}, \ldots,v_{Q^{(j)},n^{(j)}}$, one for each of the $Q$ functions computed in job $j$. We assume that all the intermediate values across the $J$ jobs have the same size of $T$ bits, which is the case when for example, we are training $J$ image classifiers in parallel using the same deep neural network.

\noindent {\bf Shuffle phase.} Before the Shuffle phase starts, for each computation job $j$, we assign the tasks of reducing the output functions symmetrically across the nodes, such that each node computes a disjoint subset of $\frac{Q}{K}$ functions. We denote the set of the indices of the output functions assigned to Node~$k$ for job $j$ as ${\cal S}_k^{(j)}$, $j=1,2,\ldots,J$.

In the Shuffle phase, 
each node~$k$ produces a message, denoted by $X_k \in \mathbb{F}_{2^{\ell_k}}$, as a function of the locally computed intermediate values in the Map phase (i.e., $\underset{n^{(j)} \in {\cal M}_k}{\cup}\{v_{1^{(j)},n^{(j)}},v_{2^{(j)},n^{(j)}},\ldots,v_{Q^{(j)},n^{(j)}}\}$), where $\ell_k \in \mathbb{N}$ denotes the length of the message in bits. Having generated $X_k$, Node~$k$ broadcasts it to all the other nodes. 

\begin{definition}[Communication Load]
We define the \emph{communication load}, denoted by $L$, as the total number of bits contained in all broadcast messages, normalized by $JQT$, i.e.,
\begin{align}
    L \triangleq \tfrac{\ell_1 + \cdots + \ell_K}{JQT}.
\end{align}
\end{definition}


\noindent {\bf Reduce phase.} For each job $j$ and each $q^{(j)} \in {\cal S}_k^{(j)}$, $j=1,2,\ldots,J$, Node~$k$ computes the output function $\phi_{q^{(j)}}$ as in (\ref{eq:linearReduce}), using the locally computed Map results and the received broadcast messages in the Shuffle phase.



\subsection{Main Results}
For the above formulated distributed computing problem, we first study the effects of applying the compression scheme and the CDC scheme individually on reducing the communication load. Then, we present our main result, which is a communication load achieved by the proposed computing scheme that jointly utilizes compression and CDC. 





Exploiting the compression technique, each sender node pre-combines all the intermediate values needed at the receiver node for a particular function, and then sends the pre-combined value. 
We demonstrate in the appendix
that the following communication load can be achieved by solely applying compression. 
\begin{align}\label{eq:sourceOnly}
 L_{\textup{compression}} = \begin{cases}
 \lceil\frac{1}{\mu}\rceil-1, &\frac{1}{K} \leq \mu <\frac{1}{2},\\
 1, &  \frac{1}{2} \leq \mu <1.
\end{cases} 
\end{align}


The above communication load achieved by compression only depends on the storage size $\mu$. In the regime of $\frac{1}{2} \leq \mu <1$, the communication load $L_{\textup{compression}}$ is a constant that does not decrease as the storage size increases. This is because that as long as $\mu <1$, each node has to receive at least one intermediate value for each of the functions it is computing.

When only applying the CDC scheme without compression, as shown in~\cite{li2016fundamental}, 
we can achieve the communication load
\begin{align}
L_{\textup{CDC}} &= \frac{(1- \mu)N}{\mu K}. \label{eq:commCDC}
\end{align}

The CDC scheme creates coded multicast packets that are simultaneously useful for $\mu K$ nodes. Hence, for fixed storage size $\mu$, the achieved communication load $L_{\textup{CDC}}$ decreases inversely proportionally with the network size ($K$). On the other hand, since the CDC scheme was designed to handle general Reduce functions that require each of the $N$ intermediate values separately as the inputs, the load $L_{\textup{CDC}}$ also scales linearly with the number of input files ($N$).





We propose the compressed coded distributed computing (compressed CDC) scheme, which jointly utilizes the combining and the coded multicasting techniques, and achieves a smaller communication load than those achieved by applying each of the two techniques individually. We present the performance of compressed CDC in the following theorem.

\begin{theorem}
To execute $J$ computation jobs with linear aggregation of intermediate results, each of which processes $N$ input files to compute $Q$ output functions, distributedly over $K$ computing nodes each with a local storage of size $\mu$, the proposed compressed CDC scheme achieves the following communication load
  \begin{align}
      L_{\textup{compressed CDC}} &=\frac{(1-\mu)(\mu K +1)}{\mu K},
  \end{align}
for $\mu K \in \{1,\ldots,K-1\}$, and $J = \gamma {K \choose \mu K+1}$, for some $\gamma \in \mathbb{N}$.
\end{theorem}

We describe the general compressed CDC scheme in the next section.

\begin{remark}
Compared with the compression scheme whose communication load is in (\ref{eq:sourceOnly}), for large $K$, the proposed compressed CDC scheme reduces the communication load by a factor of $\mu$ when $\frac{1}{K} \leq \mu < \frac{1}{2}$, and by a factor of $1-\mu$ when $\frac{1}{2} \leq \mu < 1$. In the scenarios where the cluster consists of many low-end computing nodes with small storage size (e.g., $\mu = \frac{1}{K}$), this bandwidth reduction can scale with the network size.
Also, in contrast to the compression scheme, the load $L_{\textup{compressed CDC}}$ keeps decreasing as the storage size $\mu$ increases. $\hfill \square$
\end{remark}

\begin{remark}
Unlike the communication load in (\ref{eq:commCDC}) achieved by the CDC scheme, the communication load achieved by the compressed CDC scheme does not grow with the number of input files. This is accomplished by incorporating the compression technique, i.e., pre-combining multiple intermediate values of the same Reduce function.
$\hfill \square$
\end{remark}



\begin{remark}
The file placement of the compressed CDC scheme is performed such that all $N$ input files of each particular computation job are placed exclusively on a unique subset of $\mu K+1$ nodes, following a repetitive pattern specified by the CDC scheme. As a result, the compressed CDC scheme executes a batch of ${K \choose \mu K+1}$ jobs in parallel.
In the Shuffle phase of compressed CDC, each computing node first pre-combines several intermediate values of a single function reduced at another node, and then applies bit-wise XOR operations on multiple such pre-combined packets to generate a coded multicast packet that is simultaneously useful for computing $\mu K$ functions. We note that these $\mu K$ functions can be different functions in the same job, as well as different functions in different jobs. 
$\hfill \square$
\end{remark}

\section{Description of the compressed CDC scheme}
In this section, we describe the proposed compressed CDC scheme, and analyze its communication load. 

We consider the storage size $\mu$ such that $\mu K \in \{1,2,\ldots,K-1\}$, and take sufficiently many computation jobs to process in parallel, where the number of jobs $J = \gamma {K \choose \mu K+1}$, for some $\gamma \in \mathbb{N}$. The proposed compressed CDC scheme operates on a batch of ${K \choose \mu K+1}$ jobs at a time, and repeats the same operations $\gamma$ times to process all the jobs. Therefore, it is sufficient to describe the scheme for the case of $\gamma=1$.

Along the general description of the compressed CDC scheme, we consider the following illustrative example.

\noindent {\bf Example (compressed CDC).} We have a distributed computing cluster that consists of $K=4$ nodes each with a storage size of $\mu = \frac{1}{2}$. On this cluster, we need to execute $J={K \choose \mu K+1}=4$ MapReduce jobs with linear Reduce functions, each of which requires processing $N = 6$ files to compute $Q=4$ output functions. Each node is responsible for computing one output function, for each of the $4$ jobs. In particular, Node~$k$ computes 
\begin{align}
    \phi_{k^{(j)}} = v_{k^{(j)},1^{(j)}}+v_{k^{(j)},2^{(j)}}+\dots+v_{k^{(j)},6^{(j)}},\label{eq:aggregate}
\end{align}
for all $j=1,\ldots,4$, where $v_{k^{(j)},n^{(j)}}$ is the intermediate value of the function $\phi_{k^{(j)}}$ of job $j$ mapped from the input file $w_{n^{(j)}}$ of job $j$. $\hfill \square$

\subsection{File placement}
For each job $j$, $j=1,2,\ldots,{K \choose \mu K+1}$, all of its input files $w_{1^{(j)}}, w_{2^{(j)}},\ldots,w_{N^{(j)}}$ are stored exclusively on a unique subset of $\mu K +1$ nodes, and we denote the set of indices of these nodes as ${\cal K}_j$. Within ${\cal K}_j$, each file $w_{n^{(j)}}$ of job $j$ is repeatedly stored on $\mu K$ nodes. In particular, we first evenly partition the files $w_{1^{(j)}}, w_{2^{(j)}},\ldots,w_{N^{(j)}}$ into $\mu K + 1$ batches, and label each batch by a unique size-$\mu K$ subset of ${\cal K}_j$, denoted by ${\cal P}_j$. Then, we store all the files in a batch on each of the $\mu K$ nodes whose index is in the corresponding subset ${\cal P}_j$. We denote the set of indices of the files from job $j$ in a batch labelled by a subset ${\cal P}_j$ as ${\cal B}_{{\cal P}_j}$. The file placement is performed such that for each ${\cal P}_j \subset {\cal K}_j$ with $|{\cal P}_j |= \mu K$, and each $n^{(j)} \in {\cal B}_{{\cal P}_j}$, we have
\begin{align}
    n^{(j)} \in {\cal M}_k,
\end{align}
for all $k \in {\cal P}_j$, where ${\cal M}_k$ is the set of indices of all files stored at Node~$k$.

Applying the above file placement, each node in ${\cal K}_j$ stores $\mu K \times \frac{N}{\mu K+1}$ files. Since each node is in ${K-1 \choose \mu K}$ subsets of $\{1,2,\ldots,K\}$ of size $\mu K+1$, it stores overall $\frac{\mu K N}{\mu K +1} \times {K-1 \choose \mu K} = \mu J N$ files, satisfying its local storage constraint.

\begin{figure}[htbp]
  \centering
  \includegraphics[width=0.35\textwidth]{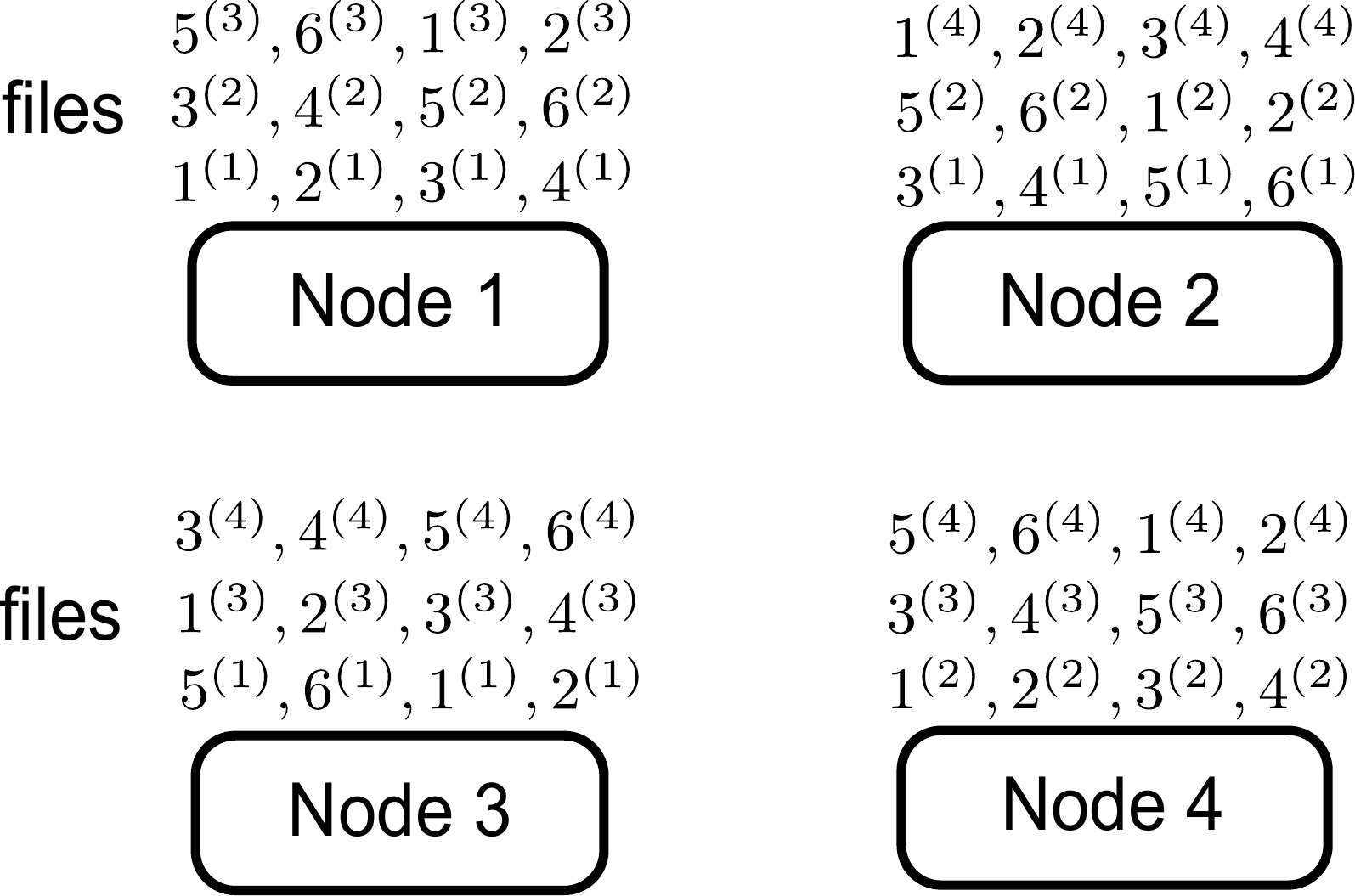}
  \caption{File placement onto $K=4$ computing nodes. For each $j=1,2,3,4$, we place the set of files for job $j$, $\{1^{(j)},2^{(j)},\ldots,6^{(j)}\}$ onto a unique subset of $\mu K+1=3$ nodes, following a repetitive pattern where each file is stored on $\mu K=2$ nodes.}
  \label{fig:CDC-partial}
\end{figure}

\noindent {\bf Example (compressed CDC: file placement).} As shown in Fig.~\ref{fig:CDC-partial}, we perform the file placement such that for each $j=1,2,3,4$, the set of files from job $j$, $\{1^{(j)},2^{(j)},\ldots,6^{(j)}\}$ are placed on a unique subset of $\mu K + 1=3$ nodes. For example, the files of job 1, $\{1^{(1)},2^{(1)},\ldots,6^{(1)}\}$ are exclusively stored on Nodes $1$, $2$, and $3$. These files are partitioned into 3 batches, i.e., ${\cal B}_{\{1,2\}} = \{3^{(1)},4^{(1)}\}$, ${\cal B}_{\{1,3\}} = \{1^{(1)},2^{(1)}\}$, and ${\cal B}_{\{2,3\}} = \{5^{(1)},6^{(1)}\}$. Then, the files $3^{(1)}$ and $4^{(1)}$ are stored on Nodes~1 and 2, the files $1^{(1)}$ and $2^{(1)}$ are stored on Nodes~1 and 3, and the files $5^{(1)}$ and $6^{(1)}$ are stored on Nodes~2 and 3. $\hfill \square$


\subsection{Coded computing}
After the file placement, the compressed CDC scheme starts the computation and data shuffling in subsets of $\mu K+1$ nodes. Within each subset ${\cal K}_j$, $j=1,2,\ldots,{K \choose \mu K+1}$, that contains the indices of $|{\cal K}_j|\!\!=\!\mu K\!+1$ nodes, the computing scheme proceeds in two stages. In the first stage, the nodes in ${\cal K}_j$ process the files they have exclusively stored, i.e., the files of job $j$. In the second stage, they handle the files from other jobs. 

\subsubsection{Stage 1 (coding for a single job)} In the first stage, nodes in ${\cal K}_j$ only process input files and compute output functions for job $j$. For ease of exposition, we drop all the job indices in the rest of the description of stage 1. According to the file placement, each node in ${\cal K}$ stores $\frac{\mu K N}{\mu K+1}$ files of job $j$, and each node in the subset ${\cal P}$ of $\mu K$ nodes stores all the files in the batch ${\cal B}_{{\cal P}}$. 

In the Map phase, each node $k \in {\cal K}$ maps all the files of job $j$ it has stored locally, for all output functions of job $j$. We note that after the Map phase, for each subset ${\cal P}$ of size $\mu K$, and $k'\in {\cal K} \backslash {\cal P}$, each of the nodes in ${\cal P}$ has computed $\frac{Q}{K}$ intermediate values, one for each of the functions assigned to Node~$k'$, from each of the files in the batch ${\cal B}_{\cal P}$. More precisely, these intermediate values are
\begin{align}
    \{v_{q,n}: q \in {\cal S}_{k'}, n \in {\cal B}_{\cal P}\}.\label{eq:intermediate}
\end{align}



In the Shuffle phase, within each subset ${\cal P} \subset {\cal K}$ of size $\mu K$, we first perform the pre-combining operation as follows. For each $k \in {\cal P}$, Node~$k$ sums up the intermediate values computed in (\ref{eq:intermediate}) to obtain the pre-combined values
\begin{align}
    \bar{v}_{q,{\cal P}} = \sum_{n\in {\cal B}_{\cal P}}v_{q,n},
\end{align}
for all $q \in {\cal S}_{k'}$.

Having computed $\frac{Q}{K}$ such pre-combined values $\{\bar{v}_{q,{\cal P}}: q \in {\cal S}_{k'}\}$, the nodes in ${\cal P}$ concatenate them to generate a packet $V_{{\cal P}}$, and evenly and arbitrarily split it into $\mu K$ segments. We label the segments by the elements in ${\cal P}$. That is, for ${\cal P} = \{i_1,i_2,\ldots,i_{\mu K}\}$, we have
\begin{align}
    V_{\cal P} = (V_{{\cal P},i_1},V_{{\cal P},i_2},\ldots,V_{{\cal P},i_{\mu K}}).
\end{align}

Finally, each node $k$ in ${\cal K}$ generates a coded packet $X_k^{\textup{stage 1}}$ by computing bit-wise XOR (denoted by $\oplus$) of the data segments labelled by $k$, i.e., 
\begin{align}
    X_k^{\textup{stage 1}} = \underset{{\cal P} \subset {\cal K}:|{\cal P}|=\mu K, k \in {\cal P}}{\oplus} V_{{\cal P},k}, \label{eq:message-stage1}
\end{align}
and multicasts $X_k$ to all other nodes in ${\cal K}$.

After Node~$k$ receives a coded packet $X_{k'}^{\textup{stage 1}}$ from Node~$k'$, it cancels all the  segments $V_{{\cal P},k'}$s with $k \in {\cal P}$, and recovers the intended segment $V_{{\cal K}\backslash \{k\},k'}$. Repeating this decoding process for all received coded packets, Node~$k$ recovers $V_{{\cal K}\backslash \{k\}}$, and hence $\bar{v}_{q,{\cal K}\backslash \{k\}}$, for all $q \in {\cal S}_k$. Using these values, together with the local Map results, Node~$k$ computes the output $\phi_{q}$ for all $q \in {\cal S}_k$. After the first stage of computation, each node in ${\cal K}_j$ completes its computation tasks for job $j$.

Since each of the coded packets in (\ref{eq:message-stage1}) contains $\frac{Q}{K} \times \frac{T}{\mu K}$ bits, the communication load exerted in the Shuffle phase of the first stage is
\begin{align}
   L_{\textup{stage 1}}=\tfrac{\tfrac{Q}{K} \times\tfrac{(\mu K+1)T}{\mu K}}{JQT} =\tfrac{ \tfrac{\mu K+1}{\mu K}}{JK}.
\end{align}

\noindent {\bf Example (compressed CDC: coding for a single job).} We start describing the proposed scheme in the subset of Nodes 1, 2, and 3. In the first stage of computation, since $\{1,2,3\}={\cal K}_1$, these three nodes will focus on processing job $1$. The computation and communication scheme for this stage is the same as described for the example in Fig.~\ref{fig:motivating-example}(c). By the end of this stage, Nodes~1, 2, and 3 compute their assigned functions for job 1. The first stage incurs 
a communication load of $L_{\textup{stage 1}}=\frac{3/2}{16}=\frac{3}{32}$. $\hfill \square$

\subsubsection{Stage 2 (coding across jobs)}
In the second stage, we first take a node $i$ outside ${\cal K}_j$, and then for each $k \in {\cal K}_j$, we label the job whose input files are exclusively stored on the nodes in $\{i\} \cup {\cal K}_j\backslash \{k\}$ as $j_k$. Next, the nodes in ${\cal P}_{j_k}={\cal K}_j\backslash \{k\}$ process the files of job $j_k$ in the batch ${\cal B}_{{\cal P}_{j_k}}$ in the Map phase, and communicate the computed intermediate values needed by Node~$k$ in a coded manner.

For a node $i \in \{1,2,\ldots,K\} \backslash {\cal K}_j$, and each $k \in {\cal K}_j$, the nodes in ${\cal P}_{j_k} = {\cal K}_j \backslash \{k\}$ share a batch of $\frac{N}{\mu K+1}$ files in ${\cal B}_{{\cal P}_{j_k}}$ for job $j_k$. In the Map phase, for each $k' \in {\cal P}_{j_k}$, Node~$k'$ computes $\frac{Q}{K}$ intermediate values, one for each function of job $j_k$ assigned to Node~$k$ in ${\cal S}^{(j_k)}_k$, from each of the files in the batch ${\cal B}_{{\cal P}_{j_k}}$. More precisely, each Node~$k'$ computes the intermediate values
\begin{align}
    \{v_{q^{(j_k)},n^{(j_k)}}:q^{(j_k)} \in {\cal S}_k^{(j_k)}, n^{(j_k)} \in {\cal B}_{{\cal P}_{j_k}}\}.\label{eq:intermediate2}
\end{align}


In the Shuffle phase, for each $k \in {\cal K}_j$, the nodes in ${\cal P}_{j_k}$ first pre-combine the Map results in (\ref{eq:intermediate2}) locally to compute
\begin{align}
    \bar{v}_{q^{(j_k)},{\cal P}_{j_k}} = \sum_{n^{(j_k)}\in {\cal B}_{{\cal P}_{j_k}}}v_{q^{(j_k)},n^{(j_k)}},
\end{align}
for all $q^{(j_k)} \in {\cal S}_k^{(j_k)}$.

Next, as similarly done in the first stage, the nodes in ${\cal P}_{j_k}$ first concatenate the above $\frac{Q}{K}$ pre-combined values $\{\bar{v}_{q^{(j_k)},{\cal P}_{j_k}}: q^{(j_k)} \in {\cal S}_k^{(j_k)}\}$ to form a packet $V_{{\cal P}_{j_k}}$, and then split it into $\mu K$ segments. We label these segments by the elements in ${\cal P}_{j_k}$, i.e., for ${\cal P}_{j_k} = \{i_1,i_2,\ldots,i_{\mu K}\}$, we have
\begin{align}
    V_{{\cal P}_{j_k}} = (V_{{\cal P}_{j_k},i_1},V_{{\cal P}_{j_k},i_2},\ldots,V_{{\cal P}_{j_k},i_{\mu K}}).
\end{align}

Finally, each node $k'$ in ${\cal K}_j$ generates a coded packet $X_{k'}^{\textup{stage 2}}$ by computing bit-wise XOR of the data segments labelled by $k'$, i.e., 
\begin{align}
    X_{k'}^{\textup{stage 2}} = \underset{t \in {\cal K}_j \backslash \{k'\}}{\oplus} V_{{\cal P}_{j_t},k'},
\end{align}
and multicasts $X_{k'}^{\textup{stage 2}}$ to all other nodes in ${\cal K}_j$.

We note that since the job index $j_t$ (whose input files are exclusively stored on nodes in $\{i\} \cup {\cal K}_j\backslash \{t\}$) is different for different $t$, the above coded packet is generated using intermediate values from different jobs.

Having received a coded packet $X_{k'}^{\textup{stage 2}}$ from Node~$k'$, Node~$k$ cancels all the segments $V_{{\cal P}_{j_t},k'}$s with $k \in {\cal P}_{j_t}$, and recovers the intended segment $V_{{\cal P}_{j_k},k'}$. Repeating this decoding process for all received coded packets, Node~$k$ recovers $V_{{\cal P}_{j_k}}$, and hence $\bar{v}_{q^{(j_k)},{\cal P}_{j_k}}$, for all $q^{(j_k)} \in {\cal S}_k^{(j_k)}$. 

We repeat the above Map and Shuffle phase operations for all $i \in \{1,2,\ldots,K\} \backslash {\cal K}_j$. By the end of the second stage, each node in ${\cal K}_j$ recovers partial sums to compute functions from $K-\mu K-1$ jobs. 

The communication load incurred in the Shuffle phase, for a particular $i$, is $\frac{\frac{Q}{K} \times\frac{\mu K+1}{\mu K}}{JQ}$, and the total communication load of the second stage is
\begin{align}
    L_{\textup{stage 2}} &=\tfrac{ (K-\mu K-1)\tfrac{\mu K+1}{\mu K}}{JK}.
\end{align}


\begin{figure}[htbp]
  \centering
  \includegraphics[width=0.48\textwidth]{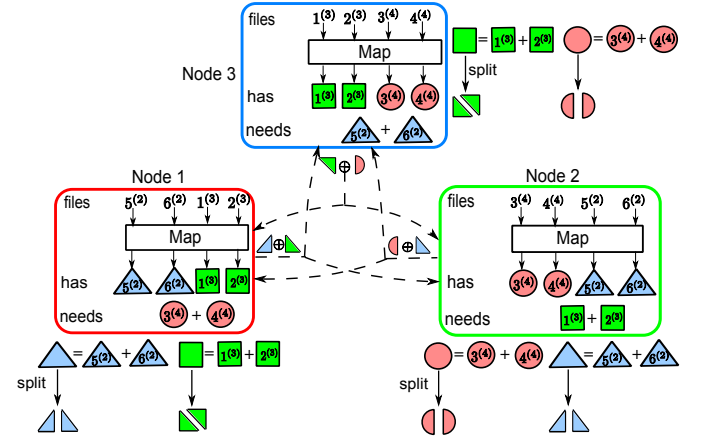}
  \caption{Illustration of the operations in the second stage of compressed CDC, in the subset of Nodes 1, 2, and 3. Note that in this stage, pre-combined packets from different jobs are utilized to create coded multicast packets.}
  \label{fig:linear-joint-out}
\end{figure}

\noindent {\bf Example (compressed CDC: coding across jobs).} We now move on to describe the second stage of compressed CDC within the subset ${\cal K}_1=\{1,2,3\}$ via Fig.~\ref{fig:linear-joint-out}, where we represent the functions computed by Node 1, 2, and 3 by red/circle, green/square, and blue/triangle respectively, and the intermediate value of a function from a file $n^{(j)}$ as the corresponding color/shape labelled by $n^{(j)}$. In this stage, as shown in Fig.~\ref{fig:linear-joint-out}, each node maps $4$ files, two of which belong to a job, and the other two belong to another job. For example, Node 1 maps the files $5^{(2)}$, $6^{(2)}$ from job $2$, and files $1^{(3)}$, $2^{(3)}$ from job $3$, producing two blue triangles labelled by $5^{(2)}$ and $6^{(2)}$, and two green squares labelled by $1^{(3)}$ and $2^{(3)}$.
During data shuffling, each node first sums up the two intermediate values from the same job to create two pre-combined packets locally (e.g., the summation of blue triangles labelled by $5^{(2)}$ and $6^{(2)}$, and the summation of green squares labelled by $1^{(3)}$ and $2^{(3)}$ at Node~$1$). Then, as shown in Fig.~\ref{fig:linear-joint-out}, each node splits each of the computed sums evenly into two segments, computes the bit-wise XOR of two segments, one from each sum, and multicasts it to the other two nodes. Finally, each node decodes the intended sum from the multicast packets using its locally computed intermediate values. The second stage incurs 
a communication load of $L_{\textup{stage 2}}=\frac{3/2}{16}=\frac{3}{32}$. $\hfill \square$

Having performed this two-stage operation on all subsets ${\cal K}_j$ of $\mu K+1$ nodes, $j=1,2,\ldots, {K \choose \mu K+1}$, each node $k$ has finished computing its assigned functions from ${K-1 \choose \mu K}$ jobs. For each of the remaining ${K \choose \mu K+1}-{K-1 \choose \mu K}$ jobs, say job $j'$, and each $k' \in {\cal K}_{j'}$, Node~$k$ receives a partial sum of $\frac{N}{\mu K+1}$ intermediate values for each of the functions in ${\cal S}_{k}^{(j')}$, in the subset $\{k\} \cup {\cal K}_{j'} \backslash \{k'\}$. Summing up these $\mu K+1$ partial sums, Node~$k$ finishes computing each of its assigned functions from job $j'$.  

The overall communication load of compressed CDC is
\begin{align}
 L_{\textup{compressed CDC}} &={K \choose \mu K+1} \!\times\! (L_{\textup{stage 1}} + L_{\textup{stage 2}}) \nonumber\\
 &=\frac{(1-\mu)(\mu K +1)}{\mu K}.
\end{align}

\noindent {\bf Example (compressed CDC: final reduction).} After the two-stage computations in the subset $\{1,2,3\}$, we repeat the same operations in the other subsets of 3 nodes. In the end, taking Node~1 as an example,
\begin{itemize}
    \item In subset $\{1,2,3\}$, Node~1 computes $\phi_{1^{(1)}}$, and $v_{1^{(4)},3^{(4)}}+v_{1^{(4)},4^{(4)}}$,
    \item In subset $\{1,2,4\}$, Node~1 computes $\phi_{1^{(2)}}$, and $v_{1^{(4)},1^{(4)}}+v_{1^{(4)},2^{(4)}}$,
    \item In subset $\{1,3,4\}$, Node~1 computes $\phi_{1^{(3)}}$, and  $v_{1^{(4)},5^{(4)}}+v_{1^{(4)},6^{(4)}}$.
\end{itemize}
Finally, Node~1 computes $\phi_{1^{(4)}}$ by adding up the received partial sums in the 3 subsets. 
We can verify that Nodes~2, 3, and 4 also successfully recover their assigned functions from the 4 jobs. The overall 
communication load is  $L_{\textup{compressed CDC}} = \frac{3}{32} \times 2 \times 4 = \frac{3}{4}$. $\hfill \square$

\begin{remark}
For the above example, using only the combining technique to process each job, we would have communicated 4 pre-combined packets, one for each node, achieving a communication load $L_{\textup{compression}} = \frac{4}{4}=1$. On the other hand, using the CDC scheme that only exploits the coded multicasting opportunities, we would have achieved a communication load of $L_{\textup{CDC}} = \frac{3}{2}$.  $\hfill \square$
\end{remark}
\section{Conclusion}
We propose a coded distributed computing scheme for MapReduce jobs with linear Reduce functions, named compressed coded distributed computing (compressed CDC), which achieves substantially smaller bandwidth consumption compared with the state-of-the-art schemes. Compressed CDC jointly exploits 1) pre-combining intermediate results for the same computation task, and 2) coded multicasting across different computation tasks, achieving significant communication reduction, compared with those achieved by applying the above two techniques separately. A future direction is to develop lower bounds on the minimum communication load, and study the optimality of the compressed CDC scheme.
\appendix[Communication Load of the compression scheme]
For the schemes that solely apply the compression/combining techniques, we consider a class of single-job strategies where we repeat the same steps to handle the scenario of executing a single job, for all $J$ jobs. Hence, it is sufficient to describe and analyze the scheme for the case where $J=1$. 
In this case, each computing node stores $\mu N$ files of a single job locally, and wants to compute $\frac{Q}{K}$ output functions.

We first consider the case of small storage size where $\frac{1}{K} \leq \mu \leq \frac{1}{2}$. In this case, we partition the indices of the input files $\{1,2,\ldots,N\}$ into $\lceil \frac{1}{\mu} \rceil$ batches, which are denoted as ${\cal B}_1,{\cal B}_2,\ldots,{\cal B}_{\lceil \frac{1}{\mu} \rceil}$. Each of the first $\lceil \frac{1}{\mu} \rceil-1$ batches contains $\mu N$ file indices, and the last batch ${\cal B}_{\lceil \frac{1}{\mu} \rceil}$ contains the remaining $N-\mu (\lceil \frac{1}{\mu} \rceil-1)N$ file indices. In the file placement phase, for each $i=1,2,\ldots,\lceil \frac{1}{\mu} \rceil$, we place the input files whose indices are in ${\cal B}_i$ in the local storage of  Nodes~$i,i+\lceil \frac{1}{\mu} \rceil,i+2\lceil \frac{1}{\mu} \rceil,\ldots$. In other words, Node $k$, $k=1,2,\ldots,K$, stores the files whose indices are in the batch ${\cal B}_{((k-1)\textup{ mod }\lceil \frac{1}{\mu} \rceil)+1}$. We note that since $\lceil \frac{1}{\mu} \rceil \leq K$, each batch of files is placed on at least one node.

In the Map phase, each node maps each of the files in the locally stored batch, generating $Q$ intermediate values for the $Q$ output functions. In the Shuffle phase, for a node $k$ to compute a function $\phi_q$ assigned to it, apart from the intermediate values computed from the local batch of files, it needs the partial sums of intermediate values from the other $\lceil \frac{1}{\mu} \rceil-1$ batches. We assume that Node~$k$ stores the files in ${\cal B}_j$ locally, then, for some other node $k'$ who stores a different batch ${\cal B}_t$, $t\neq j$, Node~$k'$ first pre-combines the intermediate values for the function $\phi_q$ to generate
\begin{align}
    \bar{v}_{q,{\cal B}_t} = \sum_{n \in {\cal B}_t} v_{q,n},
\end{align}
and sends this pre-combined package to Node~$k$. Having received $\lceil \frac{1}{\mu} \rceil-1$ such pre-combined  packets, one from a node who stores a distinct batch of files, Node~$k$ compute the function $\phi_q$ by summing them up together with the intermediate values computed from the local batch. In this communication scheme, each node receives $\lceil \frac{1}{\mu} \rceil-1$ pre-combined packets, each of which has the same size as a single intermediate value, for each of its assigned functions, incurring a total communication load of 
\begin{align}
    L_{\textup{compression}}|_{\frac{1}{K} \leq \mu \leq \frac{1}{2}} = \frac{(\lceil \frac{1}{\mu} \rceil-1)T \times \frac{Q}{K} \times K}{QT} = \lceil \tfrac{1}{\mu} \rceil-1.
\end{align}

We note that for the case of $\mu =\frac{1}{2}$, we have a total of $2$ batches, and each node only receives a single pre-combined packet to compute each of its assigned functions, resulting in a total communication load of $1$. For the cases where $\frac{1}{2} < \mu <1$, since each node has to receive at least one intermediate value to compute each of its assigned functions, the incurred communication load is at least $1$. Hence, 
increasing the storage size $\mu$ beyond $\frac{1}{2}$ does not further reduce the communication load, and we have 
\begin{align}
    L_{\textup{compression}}|_{\frac{1}{2} \leq \mu <1} = 1.
\end{align}

\bibliographystyle{IEEEtran}
\bibliography{ref}
\end{document}